\documentclass[twocolumn,showpacs,prl,floatfix]{revtex4}

\usepackage{graphicx}
\usepackage[center]{subfigure}
\usepackage{amsmath}

\begin{document}

\title{Molecular Dynamics on Diffusive Time Scales from the Phase Field Crystal Equation}

\author{$^1$Pak Yuen Chan, $^1$Nigel Goldenfeld and $^2$Jon Dantzig}
\affiliation{$^1$Department of Physics, University of Illinois at
Urbana-Champaign, Loomis Laboratory of Physics, 1110 West Green
Street, Urbana, Illinois, 61801-3080}
\affiliation{$^2$ Department of Mechanical Science and Engineering,
University of Illinois at Urbana-Champaign, 1206 West Green Street,
Urbana, Illinois, 61801-2906.}

\begin{abstract}
We extend the phase field crystal model to accommodate exact atomic
configurations and vacancies by requiring the order parameter to be
non-negative.  The resulting theory dictates the number of atoms and
describes the motion of each of them.  By solving the dynamical
equation of the model, which is a partial differential equation, we are
essentially performing molecular dynamics simulations on diffusive
time-scales.  To illustrate this approach, we calculate the two-point
correlation function of a liquid.
\end{abstract}


\pacs{02.70.Ns, 05.70.Ln} \maketitle

Molecular dynamics (MD) has long been a powerful tool to study
statistical mechanical systems (for an introduction, see (e.g.)
Ref.~\cite{rapaport2004amd}).  By postulating the interaction between
atoms and solving the resulting equations of motion, precise
information about each atom is known.  One of the drawbacks of MD,
however, is that too much information is captured.  For example, atomic
motions in MD simulations are resolved on atomic time scales, whereas in
many systems the relevant time scales are diffusive.  This makes MD
computationally demanding, if not completely inapplicable, in many
cases of interest where long time scales are required.  In this paper we
pursue a novel approach to attain long time scales, starting not from
individual particles but from a continuum description of matter known
as the phase field crystal (PFC) model\cite{BRAZ75, Elder02, Elder04,
Elder07,provatas2007upf}.

The starting point of the PFC model is that crystalline materials are
governed by a free energy functional that penalizes departures from
periodicity of the density in the same way that the Landau theory of
phase transitions uses a functional that penalizes spatial gradients of
the order parameter. The PFC model is formulated in terms of an order
parameter representing the local density, and is constructed so that
the free energy functional is minimized by a periodic order parameter
configuration.
Despite its simplicity and minimal physical input, the PFC model can
reproduce both qualitative and semi-quantitative (i.e. scaling)
properties of multicrystalline solidification\cite{Elder04}, dislocation
dynamics\cite{Berry06}, fracture, grain boundary
energetics\cite{Elder04}, elastic (phonon) interactions\cite{Stefa06},
grain coarsening\cite{Singe06}, linear and nonlinear
elasticity\cite{Elder04}, and plasticity\cite{Chan07}. The PFC model
has also been extended to binary systems\cite{Elder07,Chan07}, and can
be related to density functional theory\cite{Elder07}.  Recent
applications of the renormalization group
technique\cite{goldenfeld2005rga,goldenfeld2006rga,Badri06} and
adaptive mesh refinement have improved the computational efficiency of
the model, with resultant computational times several orders of
magnitude times faster than MD\cite{goldenfeld2005rga, athreya2007amc}.

Although the PFC model represents microscopic configurations, it is not
MD. The model describes the collective properties of
the crystal, but it does not attempt to describe the motion of each
individual atom.   One can regard the peaks in the order parameter as
representing local density maxima, and thus be identified as PFC
`atoms'.  However, although the order parameter, $\rho(\vec{x},t)$,
tends to form PFC `atoms' in order to minimize the total energy of
the system, their number is not conserved.  This neglect of the actual
atomic configuration, and the resulting absence of vacancies in the
description, prevents us from using the model to describe faithfully
microscopic phenomena that involve atomic hopping and vacancy
diffusion.

The goal of this paper is to modify the PFC model such that it  describes
not only the collective behavior, but also the motions of individual
atoms.  We will see that this can be done by constraining the value of the
order parameter to be positive.  By so doing, instead of being an abstract
order parameter, $\rho(\vec{x},t)$ becomes a physical density---the number
of atoms in the model can be controlled by adjusting a single parameter,
$\rho_0$.  The resulting theory is a MD simulation: we can specify the
temperature, number of atoms and the interaction potential between atoms.
As an illustration of this approach we simulate a simple liquid and
reproduce the form of the standard two-point pair distribution function.

\medskip \noindent {\it Inclusion of Vacancies:-\/} In real materials,
vacancies are present when the local density is low, \textit{i.e.,} when
there are not enough atoms to fill the space.  In the PFC model, however,
even if the value of the order parameter is small, which is analogous to
the low density situation, a perfect periodic configuration can still be
formed because there is no constraint, or energy penalty, for negative
values of the order parameter. Therefore, as long as the system is in a
periodic state, such as the 2-D triangular phase, any uniform
configuration will evolve to a spatially periodic one in equilibrium.
Thus, the notion of vacancies is not respected in this model.  If a
vacancy is created through a special initial condition, the free volume
will simply diffuse throughout the crystal as the configuration readjusts
its periodicity.

We can stabilize vacancies by imposing a constraint
on the order parameter---we forbid the order parameter to be negative.
In this case, if the local order parameter is not high enough, instead
of forming a periodic state that extends to negative values, the system
can form a periodic structure in some region, while leaving a very low,
or zero, density in another.  The number of atoms is then conserved and
the zero density regions are identified with vacancies.

We now identify the region of the phase diagram in which vacancies are
present and stable, and we do this by calculating the energy of a state
with vacancies, working for simplicity in two dimensions (2D). The PFC
model is given by the free energy density\cite{Elder02,Elder04},
\begin{equation}
f = \frac{\rho}{2}\left(r+(1+\nabla ^2)^2\right)\rho+\frac{\rho ^4}{4},
\label{eqn_pfc_f}
\end{equation}
where $r<0$ is the undercooling parameter and $\rho_0$ is the mean value
of the order parameter.  The dynamics follows the ``Modified'' PFC
formulation\cite{Stefa06}
\begin{equation}
\frac{\partial^2 \rho}{\partial t^2} + \beta \frac{\partial \rho}{\partial t}
= \alpha^2 \nabla^2 \frac{\delta F}{\delta \rho} + \eta
\end{equation}
where $F\equiv\int f d^d x$ is the total free energy of the system,
$\alpha$ and $\beta$ are parameters that control the evolution, and
$\eta$ is a Gaussian white noise satisfying the usual
fluctuation-dissipation theorem.
It is helpful to introduce the ansatz for the one-mode approximation to the
triangular state in two-dimensions,
\begin{equation}
\rho(\vec{x}) = A\sum_{j=1}^3 \left(e^{i\vec{k}_j\cdot\vec{x}}
               +e^{-i\vec{k}_j\cdot\vec{x}}\right) + \rho_0,
\label{eqn_tri_ansatz}
\end{equation}
where $\vec{k}_{1,2,3} = \hat{x}, (\sqrt{3}/2) \hat{y} \pm (1/2)
\hat{x}$ are the basis wavevectors of the triangular phase. Substituting
this ansatz into Eq.~(\ref{eqn_pfc_f}), and averaging
over the whole system gives the free energy density as a function of
the constant amplitude, $A$:
\begin{align}
f_0(\rho_0,A)=\frac{45}{2}A^4 &-12A^3\rho_0 +
               \frac{\rho_0^2}{4} (2+2r+\rho_0^2) \nonumber \\
          &+3A^2(r+3\rho_0^2).
\label{eq:f_0}
\end{align}
Minimizing $f_0(\rho_0,A)$ with respect to $A$ gives
two roots
\begin{equation}
A_{\pm}(\rho_0)=\frac{1}{15}\left(3\rho_0\pm \sqrt{-15r-36\rho_0^2}\right),
\label{eqn_A_solution}
\end{equation}
where the solutions that minimize the energy are $A=A_+$ for $\rho_0 >0$,
and $A=A_-$ for $\rho_0 < 0$. The roots are real for
$\rho_0 < \sqrt{-5r/12}$ (recall that $r<0$).

Now, let us consider the effect of the constraint that the density be
positive.  Examining Eq.~(\ref{eqn_tri_ansatz}), we see that the summation
is bounded by $\pm 6$, so requiring $\rho(\vec{x},t) \ge 0$ is equivalent
to requiring $|A|\le \rho_0/6$. However, Eq.~(\ref{eqn_A_solution}) shows
that $|A_+(\rho_0)|>\rho_0/6$ for all values of $r$ and $\rho_0$, so the
ground state $A=A_+$ is forbidden by the constraint. The ground state must
be given by some other configuration.

There are at least two possible configurations for the ground state.
First, the ground state can still be perfectly periodic with an
amplitude $A\neq A_+$ satisfying $|A|<\rho_0/6$.  Second, the ground
state can partition itself into two domains---a perfectly periodic
domain with average density $\rho_1$ and amplitude $A_1$ satisfying $|A_1|\le
\rho_1/6$, and a domain with $\rho(\vec{x})=0$.  The second domain
corresponds to vacancies.  To see which is realized in practice, we
have to calculate the energy of these two states, and recognize that
the ground state is the one with lower total energy.

Let us first calculate the free energy density of a perfectly triangular
state. Since $A=A_+$ is forbidden, we are left with three options for
$A$: $A=A_-$, which is the other local minimum of the free energy,
and $A=\pm\rho_0/6$. The latter two are the boundary values
satisfying the condition $|A|\le \rho_0/6$. By examining
Eq.~(\ref{eq:f_0}), one can see that
$f_0(\rho_0,\rho_0/6)\le f_0(\rho_0,-\rho_0/6)$, so we can ignore
the $A=-\rho_0/6$ solution.  The free energy density of the periodic state
is then
\begin{equation}
f_{per}(\rho_0) = f_0\left(\rho_0,\frac{\rho_0}{6}\right)
\end{equation}
if $|A_-(\rho_0)|>\rho_0/6$, and otherwise,
\begin{equation}
f_{per}(\rho_0) = \text{Min}
\left(f_0\left(\rho_0,A_-(\rho_0)\right),
  f_0\left(\rho_0,\frac{\rho_0}{6}\right)\right)
\end{equation}
where Min$(a,b)$ denotes the minimum of $a$ and $b$. Substituting
$A_-(\rho_0)$ and $\rho_0/6$ into Eq.~(\ref{eq:f_0})
gives the explicit expressions
\begin{align}
f_0(\rho_0,A_-(\rho_0)) &= -\frac{13}{500}\rho_0^4 +
  \frac{7r+25}{50}\rho_0^2 - \frac{1}{10}r^2\nonumber\\
& \quad -\frac{20r\rho_0+48\rho_0^3}{375}\sqrt{-15r-36\rho_0^2}
\label{eq:f0_A} \\
f_0\left(\rho_0,\frac{\rho_0}{6}\right) &=
   \frac{1}{288}\left[133\rho_0^4+(144+168r)\rho_0^2 \right].
   \label{eq:f0_A6}
\end{align}

Now, let us compare the energy of these two possible ground states.  If
the system is perfectly periodic over the whole domain, whose area is
designated $B_0$, then the free energy
is given by
\begin{equation}
f_{whole}(\rho_0) = B_0 f_{per}(\rho_0),
\end{equation}
If the whole system instead partitions
itself into one domain made up of a triangular phase having mean density
$\rho_1 >
\rho_0$, with the remaining domain having $\rho=0$,
the free energy is given by (for simplicity, surface energy between the
two phases is neglected in this calculation.)
\begin{equation}
f_{v}(\rho_0) = B_1f_{per}(\rho_1) = \left(\frac{\rho_0}{\rho_1}\right)B_0
  f_{per}(\rho_1),
\end{equation}
where $B_1$ is the area of the triangular domain.  The second equality is
obtained by using the conservation of mass $\rho_0B_0 = \rho_1b_1$.  The
difference between these two free energies, $\Delta f\equiv
f_{v}-f_{whole}$, is
\begin{equation}
\Delta f =B_0\rho_0\left(\frac{f_{per}(\rho_1)}{\rho_1}
         -\frac{f_{per}(\rho_0)}{\rho_0}\right).
\end{equation}
It is important to note that $\rho_1$ is a
parameter that we can choose to minimize the energy of the second possible
state; the only constraint is that $\rho_1\ge \rho_0$ because $B_1 \le
B_0$.

\begin{figure}[htb]
\begin{center}
\includegraphics[width=0.75\columnwidth]{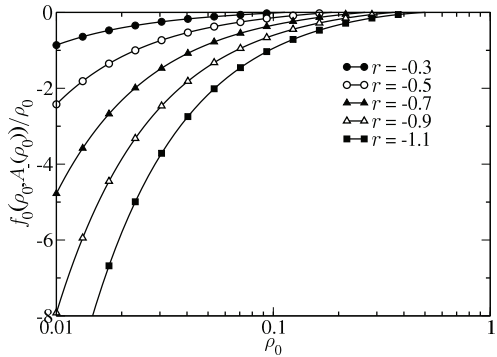}
\end{center}
\caption{The function $f_0(\rho_0,A_-(\rho_0))/\rho_0$ is an
increasing function of $\rho_0$, plotted for various values of $r$. The x-axis
is plotted on logarithmic scale in order to resolve the curves.}
\label{fig_A_energy}
\end{figure}

For vacancies to exist, we require that $\Delta f < 0$ for some values of
$\rho_1>\rho_0$.  We note, however, that for the solution $A=A_-$,
$f_0(\rho_0,A_-(\rho_0))/\rho_0$ is an increasing function of $\rho_0$
(see Fig.~(\ref{fig_A_energy})) and so $\Delta f$ is positive for this
branch of the solution.  In other words, no vacancy is present in this
solution. Therefore, in order to have vacancies in the ground state, we
require this branch of the solutions to be forbidden by the constraint;
i.e., we require $|A_-(\rho_0)|>\rho_0/6$, which by
Eq.~(\ref{eqn_A_solution}) is equivalent to requiring
\begin{equation}
\rho_0, \rho_1 < \sqrt{-{12 r}/{53}}.
\label{eqn_rho_constraint}
\end{equation}

On the other hand, it is easy to show from Eq.~(\ref{eq:f0_A6}) that
$f_0(\rho_0,\rho_0/6)/\rho_0$ has  a minimum (for $r < -6/7$) at
\begin{equation}
\rho_{min}=\sqrt{(-48-56r)/133}
\label{eq:rho_min}
\end{equation}
Thus, if $\rho_0 \le \rho_{min}$ and $r < -6/7$, the system can minimize
the total free energy by partitioning itself into two domains: a
triangular phase made up of 'atoms', with average density  $\rho =
\rho_{min}$, and a region of vacancies where $\rho=0$. Combining
Eqs.~(\ref{eqn_rho_constraint}) and (\ref{eq:rho_min}) indicates that the
minimum also satisfies the constraint $|A_-(\rho_0)|>\rho_0/6$ so long as
$r>-636/343$.  We also note that the area of the triangular phase,
$B_1=B_0(\rho_0/\rho_1)$, is directly proportional to the mean density,
$\rho_0$.  So by adjusting $\rho_0$, we can control the number of atoms in
the PFC model.  This shows that the addition of the constraint,
$\rho(\vec{x})>0$ for all $\vec{x}$, does indeed promote the
$\rho(\vec{x})$ from an abstract order parameter to a physical density,
which dictates the number of atoms in the system.

\begin{figure}[htb]
\centering
\includegraphics[width=0.75\columnwidth]{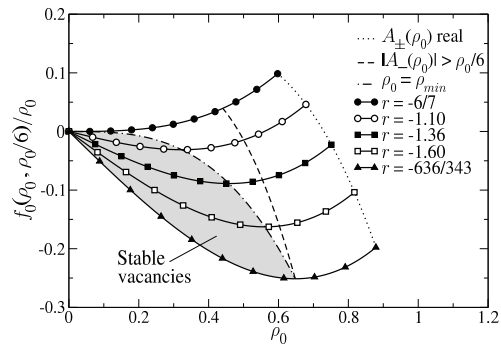}
\caption{Window in which stable vacancies can coexist with a triangular
phase. A minimum in
$f_0(\rho_0,\rho_0/6)/\rho_0$ exists at $\rho_{min}=\sqrt{(-48-56r)/133}$
for $r\le -6/7$, and this minimum is forbidden by the constraint
$|A_-(\rho_0)| \le \rho_0/6$ for $r > -636/343$.}
\label{fig:vac_window}
\end{figure}

To summarize, the various constraints define the region
\begin{equation}
\rho_{min}<\sqrt{\frac{-48-56r}{133}} \quad\text{and}\quad
-\frac{636}{343}< r <-\frac{6}{7},
\label{eqn_vac_phase}
\end{equation}
where the triangular phase and stable vacancies can coexist. The area
fraction of the triangular phase is $\rho_0/\rho_{min}$, and the amplitude
is  $A=\rho_{min}/6$. The rest of the domain has zero density and thus is
composed of vacancies. These results are summarized in
Fig.~(\ref{fig:vac_window}).

\medskip
\noindent {\it Implementation:-\/} In order to implement the positive
density constraint, we add a vacancy term, $f_{vac}(\rho)$, to the free
energy density that penalizes negative values
of $\rho(\vec{x},t)$. As long as the repulsion from negative values is
strong enough to avoid $\rho<0$, the result should not depend on any
particular choice of $f_{vac}(\rho)$. Of the many possible choices
for $f_{vac}(\rho)$, we use
\begin{equation}
f_{vac}(\rho) = H (|\rho|^n-\rho^n),
\label{eqn_vac_f}
\end{equation}
with $n=3$ and $H=1500$, because this turns out to be numerically
convenient and stable.

\begin{figure}[htb]
\begin{center}
\begin{tabular}{c}
\includegraphics[width=0.33\columnwidth]{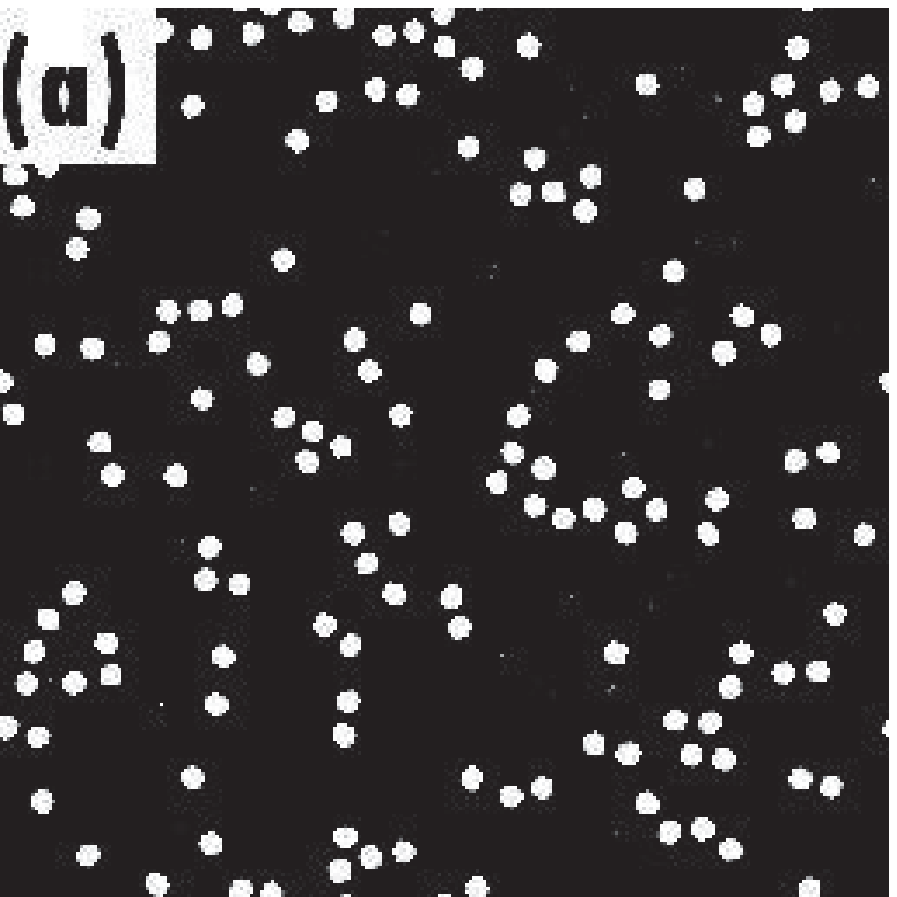}
\includegraphics[width=0.33\columnwidth]{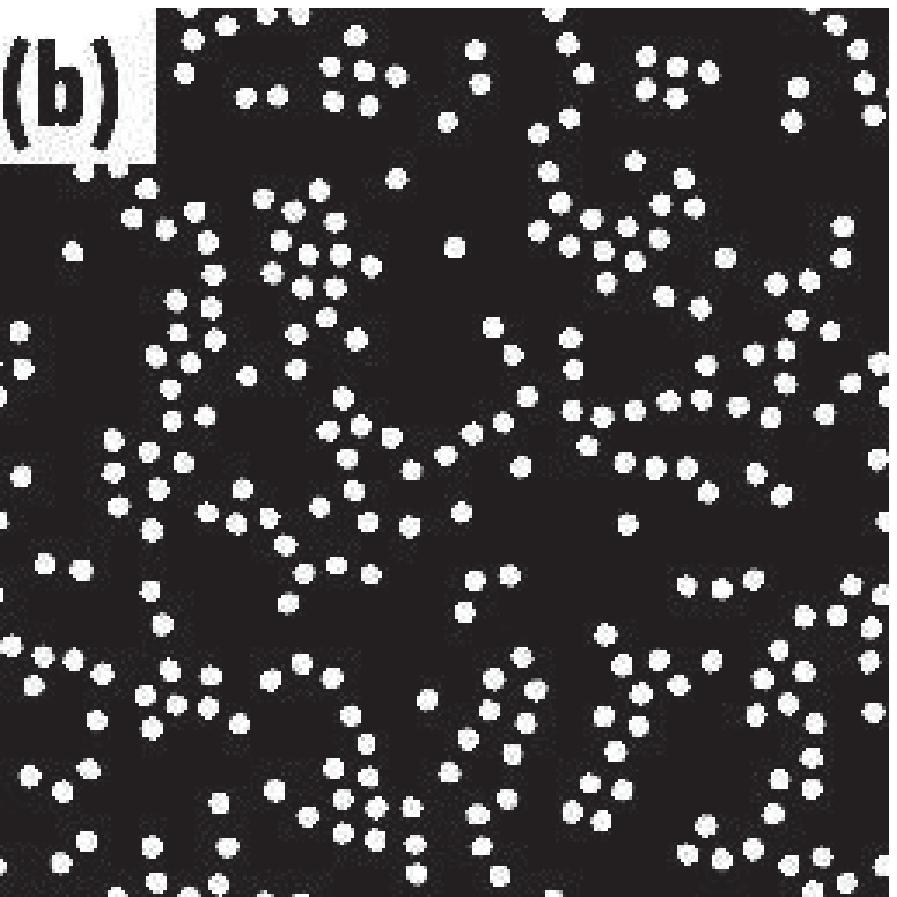}
\includegraphics[width=0.33\columnwidth]{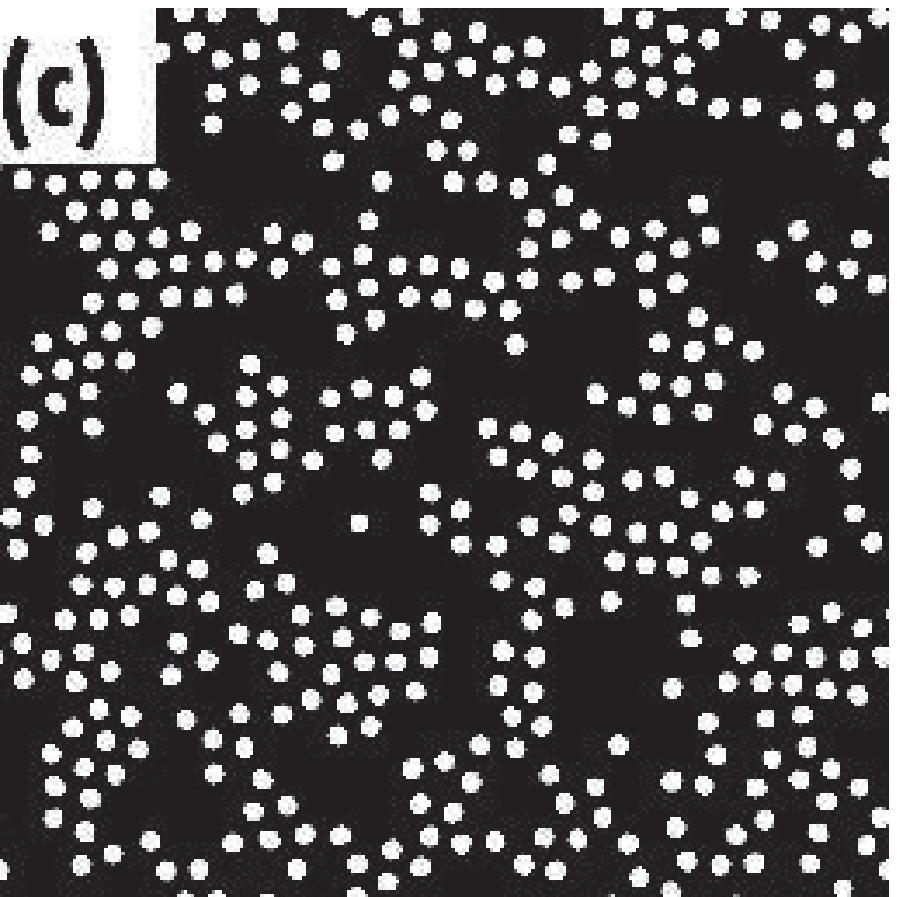}\\
\includegraphics[width=0.33\columnwidth]{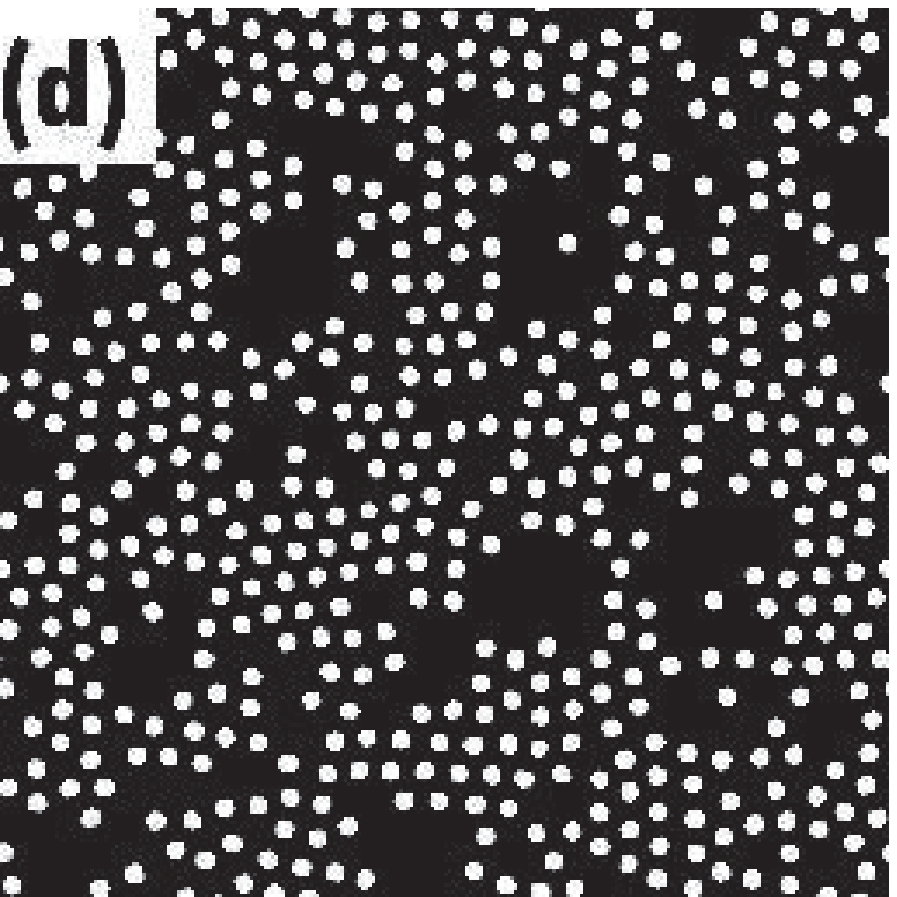}
\includegraphics[width=0.33\columnwidth]{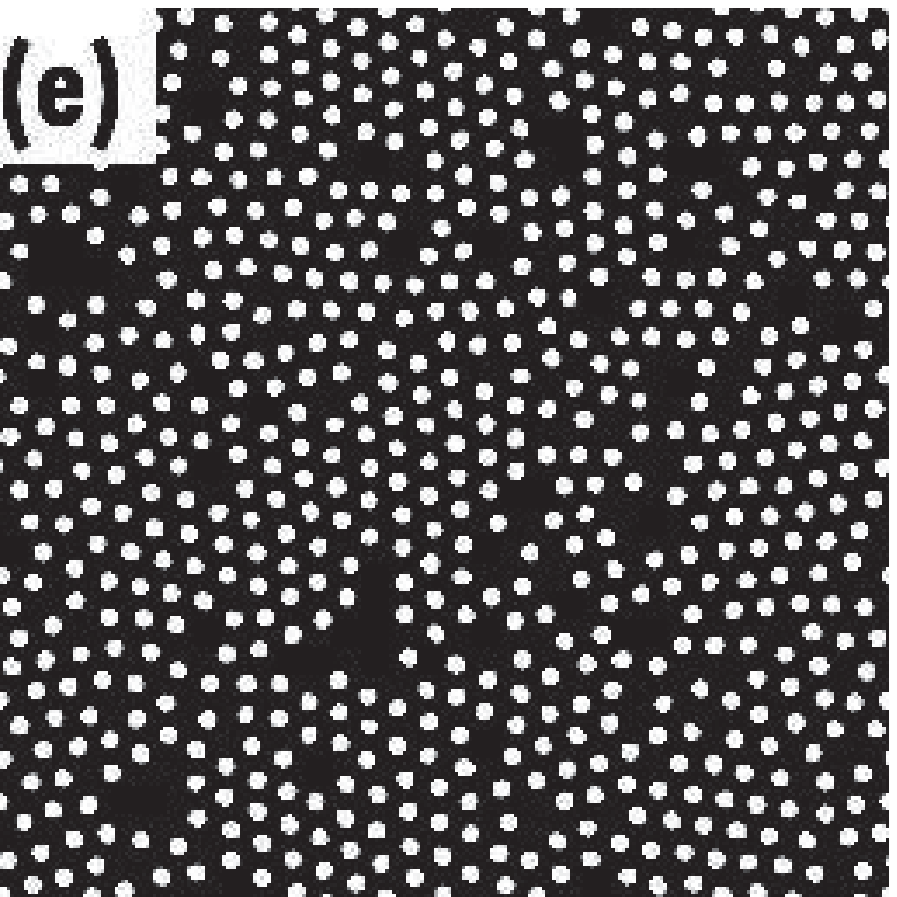}
\includegraphics[width=0.33\columnwidth]{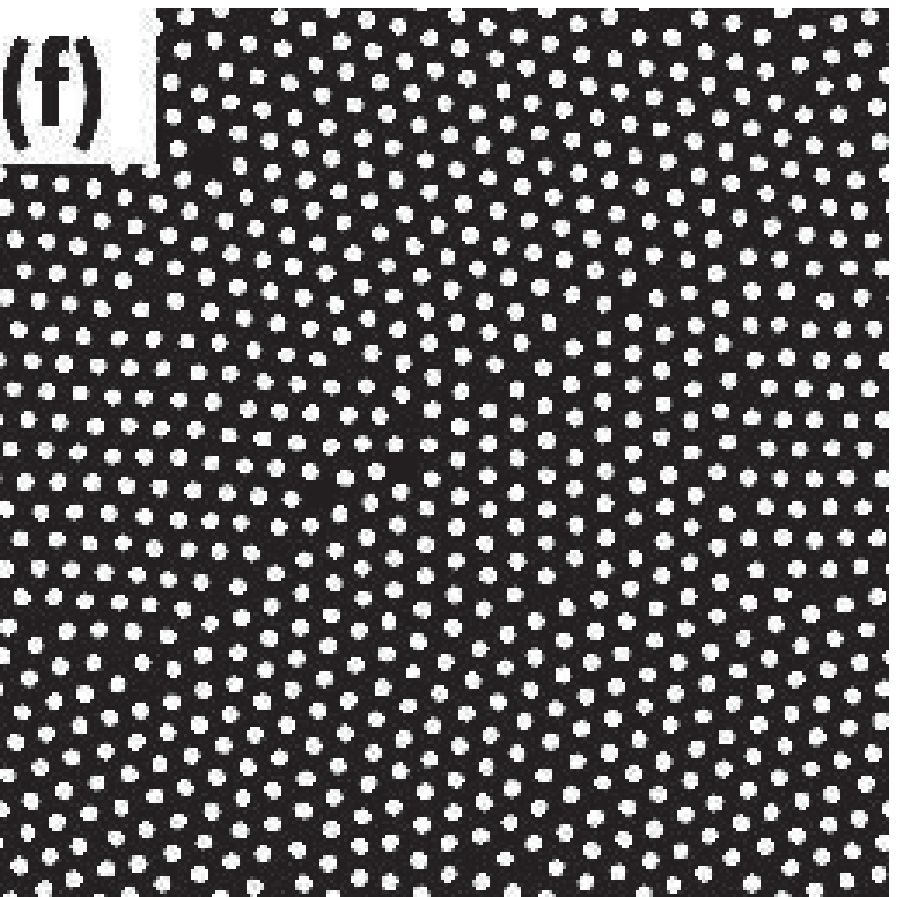}
\end{tabular}
\end{center}
\caption[Vacancies in the PFC model.] { PFC simulations with different
values of $\rho_0$ for $r=-0.9$.  The number of atoms increases with
$\rho_0$.  (a)-(f) correspond to
$\rho_0=0.06$, $0.08$, $0.10$, $0.12$, $0.14$ and $0.16$. }
\label{fig_PFC_vac}
\end{figure}

\begin{figure}[b]
\begin{center}
\includegraphics[width=0.75\columnwidth]{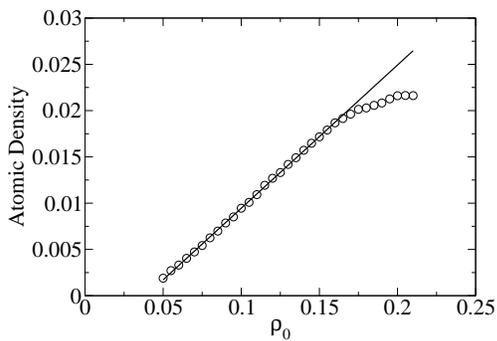}
\caption[Linear dependence of the PFC atomic density on $\rho_0$.] {
The PFC atomic density increases linearly with the order parameter,
$\rho_0$, when the vacancy term is added to the model.  $r=-0.9$ is
used. The curve starts to saturate at around $\rho_0=0.15$, as opposed
to the analytical prediction $\rho_0=\sqrt{(-48-56r)/133}=0.134$. }
\label{fig_pfc_density}
\end{center}
\end{figure}

With the vacancy term, Eq.~(\ref{eqn_vac_f}), we can numerically verify
the analytical calculation for the coexistence between the periodic phase
and vacancies.  Fig.~(\ref{fig_PFC_vac}) provides results from simulations
with $r=-0.9$ and different values of $\rho_0$, showing clearly that the
number of atoms increases with $\rho_0$.  In addition,
Fig.~(\ref{fig_pfc_density}) shows that the PFC atomic density
(\textit{i.e.,} the number of atoms per unit area.) indeed increases
linearly with $\rho_0$, as expected.   The curve starts to saturate at
around $\rho_0=0.15$, as opposed to the prediction from
Eq~(\ref{eqn_vac_phase}), $\rho_{min} \approx 0.134$.  This discrepancy is
not surprising for several reasons: In the analytical calculation, we
consider only the one-mode approximation in the ansatz; we did not
account  for the surface energy between the triangular phase and the
vacancies; and we did not account for thermal fluctuations, introduced in
the simulation to help the system equilibrate faster.

\begin{figure}[htb]
\begin{center}
\includegraphics[width=0.75\columnwidth]{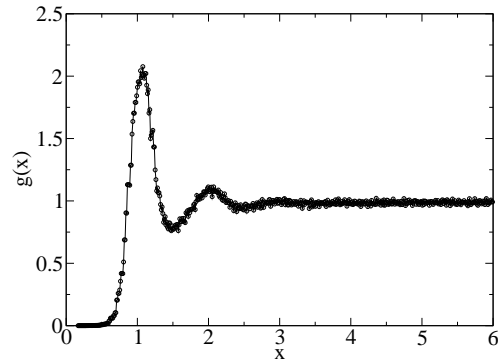}
\caption[Two point correlation function of a liquid using the PFC
model.] { The two point correlation function of a liquid using the PFC
model.  Parameters are $r=-0.9$, $\rho_0=0.09$, $\alpha=15$ and
$\beta=0.9$. } \label{fig_pfc_liq_g}
\end{center}
\vskip -0.2truein
\end{figure}

With the modifications described above,the PFC simulation operates very
much like a molecular dynamics simulation, but on diffusive time scales
many orders of magnitude faster than pure molecular
dynamics\cite{Elder04}.  We can control the number of atoms and the
temperature in the system by adjusting $\rho_0$ and the magnitude of
thermal noise, $\eta$, respectively.  The interaction potential
between individual PFC atoms is specified by the PFC free energy
(specifically the gradient terms) and is controlled by the undercooling
$r$.  In fact, by decreasing the value of $\rho_0$ such that the system is
dilute enough, we can simulate a liquid using the PFC model!  We simulated
such a liquid with parameters $r=-0.9$, $\rho_0=0.09$, $\alpha=15$ and
$\beta=0.9$. A typical result is shown in Fig.~(\ref{fig_PFC_vac}(b)).
Fig.~(\ref{fig_pfc_liq_g}) shows the two point correlation function,
$g(x)$, extracted from the simulation.  It resembles the two point
correlation function of a liquid---a correlation hole, a strong nearest
neighbor correlation and a weak correlation with atoms one or two atomic
spacings away\cite{rapaport2004amd}.

\acknowledgments{

We are grateful to Ken Elder and Nik Provatas for helpful discussions.
We acknowledge partial support from the National Science Foundation
through grant no. NSF-DMR-01-21695.}


\bibliographystyle{apsrev}

\bibliography{pfc_vacancy_bib}

\end{document}